\documentclass[twocolumn, pra, superscriptaddress]{revtex4-1}
\usepackage[T1]{fontenc}
\usepackage{lmodern}
\usepackage{textcomp}
\usepackage{amssymb}
\usepackage{amsmath}
\usepackage{epsfig}
\usepackage{color}
\usepackage{graphics, graphicx}
\usepackage{bbold}
\usepackage{psfrag}
\usepackage{mathcomp}
\usepackage[caption=false]{subfig} 
\usepackage{verbatim}
\usepackage{color}
\usepackage[colorlinks,citecolor=blue]{hyperref}

\usepackage{bm}
\usepackage{physics}
\usepackage{newtxmath}
\usepackage{mathtools} 
\usepackage{ulem} 
\def\dsi{\vmathbb{i}}

\begin{document}


\title{Experimental engineering of Floquet topological phases in a one-dimensional optical lattice}


\author{Pengju Zhao}
\affiliation{ 
  International Center for Quantum Materials, School of Physics, Peking University, Beijing 100871, China
}%
\affiliation{ 
  Beijing Aerospace Institute for Metrology and Measurement Technology, Beijing 100076, China
}%
\affiliation{
  Beijing Key Laboratory of Aerospace Photonic Quantum Precision Measurement and Metrology, Beijing 100076, China
}
\author{Yudong Wei}
\email{weiyd2017@pku.edu.cn}
\author{Zhongshu Hu}%
\author{Shengjie Jin}
\affiliation{ 
  International Center for Quantum Materials, School of Physics, Peking University, Beijing 100871, China
}%
\affiliation{ 
  Hefei National Laboratory, Hefei 230088, China
}%
\author{Xuzong Chen}
\affiliation{%
School of Electronics Engineering and Computer Science, Peking University, Beijing 100871, China
}%
\author{Xiong-jun Liu}
\email{xiongjunliu@pku.edu.cn}
\affiliation{ 
  International Center for Quantum Materials, School of Physics, Peking University, Beijing 100871, China
}%
\affiliation{ 
  Hefei National Laboratory, Hefei 230088, China
}%
\affiliation{
  International Quantum Academy, Shenzhen 518048, China
}

\date{\today}

\begin{abstract}
Periodic driving enables realization of topological phases without static counterparts. We experimentally realize and detect a one-dimensional anomalous Floquet topological phase in an optical lattice, using multi-frequency control to manipulate the relative sign structure of the gap windings $(W_0,W_\pi)$ associated with the $0$ and $\pi$ quasienergy gaps. We develop a lattice-depth modulation scheme that induces staggered nearest-neighbor $s$-$p$ orbital couplings and realize a minimal nontrivial Floquet topology under single-tone driving. 
Introducing a second tone, its relative phase controls the effective coupling signs in the $0$ and $\pi$ gaps, thereby tuning the corresponding windings to add and produce a high-winding phase or to cancel while retaining nontrivial gap indices. We read out $(W_0,W_\pi)$ with a band-inversion-surface (BIS)-resolved Ramsey protocol assisted by lattice-position shaking, which measures relative Floquet phases on the BISs. Controlled quenches further confirm phase-dependent band modifications even at quasimomenta far from resonance. These results establish multi-frequency control with a tunable relative phase as a quantitative route to engineering anomalous Floquet topology, and demonstrate phase-coherent coexistence of distinct drive modalities.
\end{abstract}

\maketitle


\section{\label{sec:level1}
introduction}

Topological phases of quantum matter have become a universal concept across modern physics. Periodic driving provides a controlled route to design such phases, including ones without static counterparts~\cite{eckardt2017,Rudner2020,Weitenberg2021}. A first theme comprises conventional Floquet topological phases, where the stroboscopic dynamics over one period can be captured by an effective static Floquet Hamiltonian $H_F$ within a high-frequency or rotating-frame description. Correspondingly, these phases obey the standard bulk-edge correspondence, essentially independent of the time dimension~\cite{PhysRevB.79.081406,PhysRevB.82.235114,rechtsman2013photonic,PhysRevX.4.031027,Jotzu2014,zheng2014,bukov2015,eckardt2015,Flaschner2016}.
A second theme is anomalous Floquet topology, in which Floquet-band invariants alone are insufficient to establish the conventional bulk-edge correspondence. Instead, topology is encoded in the full one-period evolution operator, beyond any static description. In particular, it can be specified gap by gap through the global winding numbers $(W_0,W_\pi)$ of the $0$ and $\pi$ quasienergy gaps~\cite{PhysRevX.3.031005,Delplace2014,Nathan_2015,PhysRevLett.114.106806}. Recent studies further reveal that Floquet engineering of local topological structures can realize unconventional Floquet phases whose boundary physics may not be fully captured by these global windings~\cite{zhangUnconventional2022}.

Beyond conceptual novelty, anomalous Floquet topology is attractive from an experimental perspective since it ties robust edge transport to quasienergy gaps  rather than to Floquet-band invariants~\cite{Leykam2016,Nathan2019}. More generally, a gap-resolved characterization remains effective even when Floquet-band invariants fail to predict edge transport, and it naturally motivates experimental control and readout by addressing the $0$ and $\pi$ gaps separately~\cite{Nathan_2015,PhysRevLett.114.106806}. In particular, photonic platforms have explored this regime extensively, including two-dimensional anomalous Floquet phases exhibiting chiral edge transport even when all Floquet-band Chern numbers vanish~\cite{maczewsky2017observation,mukherjee2017experimental,Chen2022,zhang2025floquet}. Related one-dimensional implementations have also observed anomalous $\pi$ modes in driven waveguide arrays~\cite{Cheng2019}. Beyond crystalline settings, engineered geometries provide additional knobs for tailoring anomalous edge transport, as exemplified by dual Sierpinski carpets~\cite{li2023fractal}.

Ultracold atoms provide a clean platform for Floquet topological matter, with tunable control of band structure, disorder, and interactions~\cite{Cooper2019,Bouchoule2025}. 
This flexibility holds promise for extending anomalous Floquet topology beyond the noninteracting limit.
Nevertheless, experimental realizations remain scarce~\cite{2020Xie,Wintersperger2020,2022Lu,Zhang2023}. To date, two key experiments have directly accessed the gap windings $(W_0,W_\pi)$. Wintersperger \textit{et al.}~\cite{Wintersperger2020} implemented an anomalous Floquet phase on a hexagonal lattice based on the step-wise modulation protocol~\cite{PhysRevB.82.235114} and inferred gap-resolved windings from energy gap measurements and local Hall deflections. Zhang \textit{et al.}~\cite{Zhang2023} realized a driven square Raman lattice and extracted windings from quench dynamics within the band-inversion-surface (BIS) framework~\cite{Zhang2020,zhangUnconventional2022}. In the latter approach, the corresponding windings can be inferred gap by gap from time-averaged spin textures across the BISs, enabling Floquet phase mapping with multiple BIS loops.
Existing cold-atom experiments have largely focused on single-tone driving, where multiple quasienergy gaps are opened via higher-order (multi-photon) processes. Systematic, phase-controlled and multi-frequency schemes for tuning individual gap windings, as well as one-dimensional anomalous Floquet phases, remain underexplored. 
Meanwhile, recent advances in orbital lattices have demonstrated coherent resonant $s$-$p$ hybridization, enabling synthetic magnetic flux~\cite{kangCreutzLadderResonantly2020} and selective control of topological pumping~\cite{Minguzzi2022,Sandholzer2022}. Beyond single-particle band engineering, orbital degrees of freedom also suggest emerging routes for tunable interactions beyond the lowest band, highlighting the broader potential of orbital platforms~\cite{Kiefer2023,Venu2023}.

Here we propose and experimentally realize one-dimensional (1D) Floquet topological phases and achieve phase-tunable control of the gap windings $(W_0,W_\pi)$ using multi-frequency lattice-depth modulation (LDM). Importantly, due to the opposite parity of the $s$ and $p$ orbitals, LDM naturally generates a first-order staggered nearest-neighbor $s$-$p$ coupling. This provides a minimal route to engineering an effective orbital-ladder Hamiltonian~\cite{Creutz1999,Sun2017,Lewenstein2017}. Previous realizations of this type of Hamiltonian typically relied on lattice-position shaking (LPS), where the corresponding effective coupling is dominated by second-order (two-photon) processes and is therefore weaker~\cite{kangCreutzLadderResonantly2020,Minguzzi2022,Sandholzer2022}.
Further, by introducing two-tone LDM at $(\omega_0,\,2\omega_0)$, we simultaneously open and address both the $0$ and $\pi$ quasienergy gaps already at first order, enabling access to both high-winding and vanishing-winding Floquet phases with sizable gaps under full control of the relative drive phase. 
To directly resolve the topology gap by gap, we implement a BIS-resolved Ramsey protocol that measures the relative phase between the two quasimomenta $k_L$ and $k_R$ associated with each gap. A robust $\pi$ phase contrast between them provides a minimal dynamical hallmark of a nontrivial winding in the corresponding gap. 

The remainder of the paper is organized as follows. Sec.~\ref{Sec2} develops the effective stroboscopic Hamiltonian relevant to the experiment and details the single-tone preparation and readout protocols, showing that LDM yields a nontrivial effective coupling between $s$-$p$ orbitals and that LPS and LDM can coherently coexist within a single sequence. Sec.~\ref{Sec3} introduces direct two-tone LDM driving. Using the same interferometric probes together with controlled quenches, we reveal phase-dependent modifications of the Floquet bands and determine the relative sign structure of the gap windings. Sec.~\ref{Sec4} summarizes the results and discusses future outlook.

\section{Experimental result}\label{Sec2}

\begin{figure*}[ht]
  \includegraphics[scale=0.23]{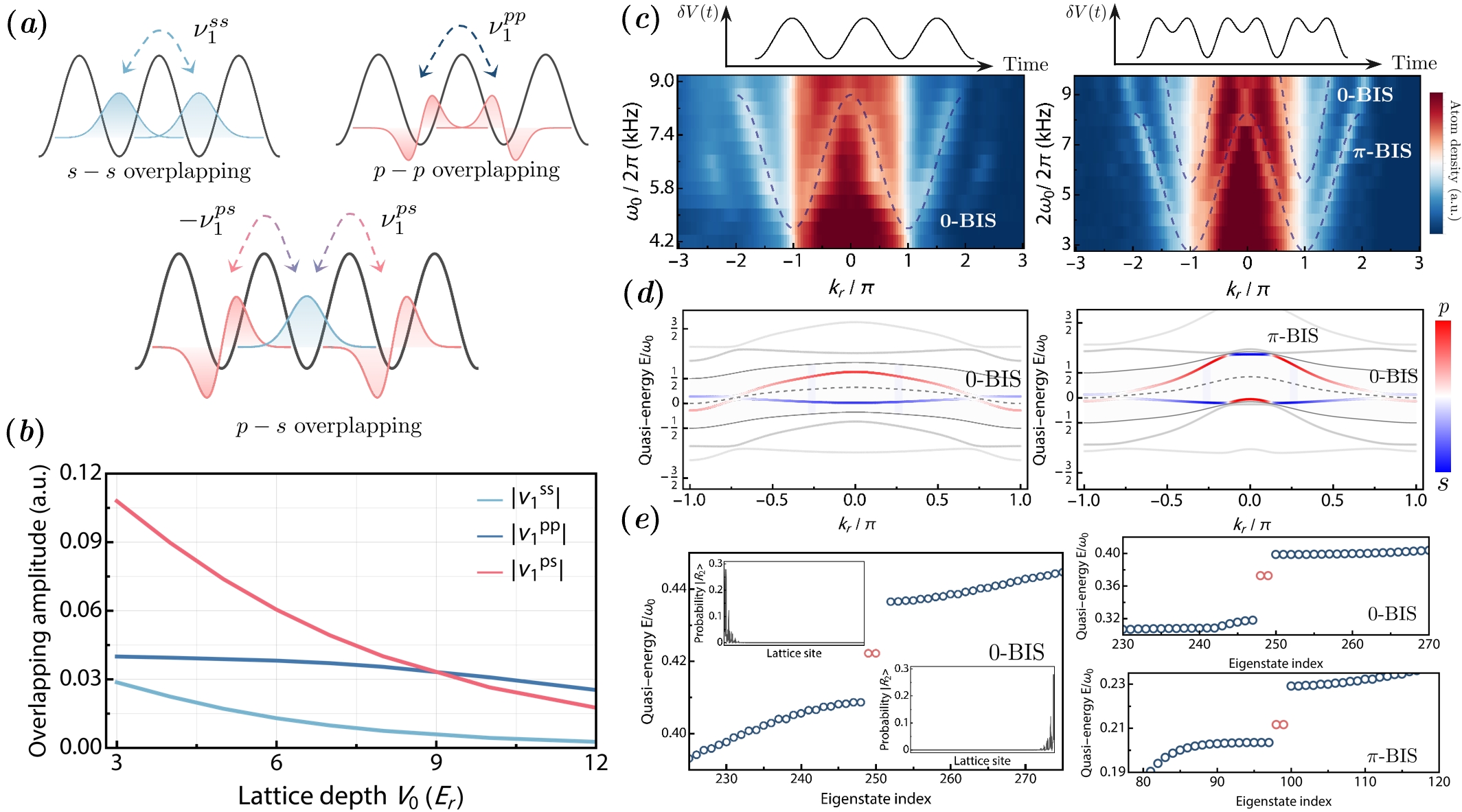}
  \caption{\label{fig1} 
Topology from lattice-depth modulation (LDM) and the corresponding Floquet band structures for single- and two-tone driving.
(a) A gauge transformation $w_p(x-x_j)\rightarrow(-1)^j w_p(x-x_j)$ renders the $p$-orbital hopping uniform while keeping the $s$-$p$ coupling staggered.
(b) Nearest-neighbor overlap coefficients $\nu_1^{\alpha\beta}$ for the $s$-$s$, $p$-$p$, and $p$-$s$ channels. The adjacent $s$ and $p$ Wannier overlap remains sizable up to $V_0=10\,E_r$.
(c) Experimental measurements are performed for single-tone driving at $\omega_0$ with modulation amplitude $\delta V_0=0.875\,E_r$ and lattice depth $V_0=4.3\,E_r$, and two-tone driving at $(\omega_0,\,2\omega_0)$ with $\delta V_0^{(1)}=\delta V_0^{(2)}=0.875\,E_r$ and lattice depth $V_0=3.5\,E_r$. All data are extracted from time-of-flight (TOF) images taken after 30 ms of expansion. Each slice is averaged over five shots.
(d,e) Numerical Floquet bands and the corresponding edge states for each BIS pair. Single-tone driving at $\omega_0=5\,E_r/\hbar$. Two-tone driving at $(\omega_0,\,2\omega_0)$ with fundamental frequency $\omega_0=4\,E_r/\hbar$. The calculation for edge states  in (e) uses $N=300$ lattice sites.}
\end{figure*}

Our experiment realizes a one-dimensional, spin-independent optical lattice described by
\begin{align}
  H(t)&=H_0+H_{\rm mod}(t),\nonumber\\
  H_0&=\frac{\hat{p}^2}{2m}+\frac{1}{2}V_0\cos(2k_L x),\nonumber\\
  H_{\rm mod}(t)&=\frac{1}{2}\delta V(t)\cos(2k_L x),
\end{align}
where $k_L=2\pi/\lambda$ with $\lambda\approx 1064\,\mathrm{nm}$. Here $V_0>0$ and $\delta V(t)$ denote the static and time-dependent lattice depths, respectively. The recoil energy is $E_r=\hbar^2 k_L^2/(2m)$, with $E_r/h \approx 2.02\,\mathrm{kHz}$. We choose the modulation to have zero time average over one period, i.e., $\langle\delta V(t)\rangle_T=0$.

Restricting to the $s$ and $p$ orbitals, we write the driven lattice in a tight-binding form,
\begin{align}\label{eq_realspaceH}
  \hat{H}(t)=\sum_i\hat{\Psi}_i^\dagger M(t)\hat{\Psi}_i+\sum_i\qty[\hat{\Psi}_i^\dagger J(t)\hat{\Psi}_{i+1}+\mathrm{h.c.}],
\end{align}
where $\hat{\Psi}_i^\dagger=(\hat{a}_{i,p}^\dagger,\hat{a}_{i,s}^\dagger)$ and $\hat{a}_{i,\alpha}^\dagger$ creates an atom in the Wannier orbital $w_\alpha(x-x_i)$ with $\alpha=s,p$.
The onsite matrix and nearest-neighbor coupling matrix are
\begin{align}
  M(t)&=\mqty(\epsilon_p&0\\0&\epsilon_s)+\mqty(\nu_0^{pp}&0\\0&\nu_0^{ss})\,\delta V(t)
\end{align}
and
\begin{align}
  J(t)&=\mqty(\tilde{t}_p&0\\0&t_s)+\mqty(\nu_1^{pp}&\nu_1^{ps}\\-\nu_1^{ps}&\nu_1^{ss})\,\delta V(t).
\end{align}
Here $\epsilon_{s/p}$ are the onsite energies for the static lattice $H_0$. The coefficients
$\nu_0^{\alpha\alpha}\equiv\frac{1}{2}\int\!dx\, w_\alpha^\ast(x-x_i)\cos(2k_L x)w_\alpha(x-x_i)$
describe the LDM-induced onsite shifts, and
$\nu_1^{\alpha\beta}\equiv\frac{1}{2}\int\!dx\, w_\alpha^\ast(x-x_i)\cos(2k_L x)w_\beta(x-x_{i+1})$
with $\alpha\beta\in\{ss,pp,ps\}$ quantify the corresponding modification of the nearest-neighbor couplings.
Because the $p$-orbital Wannier function has odd parity, the LDM-induced $s$-$p$ coupling changes sign under bond reversal, which provides the antisymmetric coupling structure required for a nontrivial band topology (Fig.~\ref{fig1}(a)(b)).

We transform to quasimomentum space via $\hat{\Psi}_k^\dagger=\frac{1}{\sqrt{N}}\sum_j e^{\dsi kj}\hat{\Psi}_j^\dagger$, giving $\hat{H}(t)=\sum_k \hat{\Psi}_k^\dagger \hat{H}_k(t)\hat{\Psi}_k$. For a minimal description, we replace the $p$-band dispersion by $\tilde{t}_p\cos k$ and fix $\tilde{t}_p$ by matching the bandwidth of a third-order cosine expansion~\cite{footnote1}. This approximation preserves the relevant gap openings and therefore does not affect the topology.
For single-tone driving $\delta V(t)=\delta V_0\cos(\omega_0 t+\phi_0)$, the time-independent effective Floquet Hamiltonian near the gap-opening takes the form~\cite{SM}
\begin{align}\label{Hamiltonian1}
H_F^{(\ell=1)}(k)=\sum_{i=y,z}h_{F,i}^{(1)}(k)\sigma_i+h_{F,0}^{(1)}(k)\sigma_0,
\end{align}
where the superscript $\ell=1$ denotes the first resonant Floquet channel. For the single-tone modulation considered here, only the first drive harmonic $\omega_0$ is present. The Hamiltonian components are
\begin{align}\label{Hamiltonian2}
  h_{F,y}^{(1)}(k)&=-\delta V_0\nu_1^{ps}\sin k,\nonumber\\
  h_{F,z}^{(1)}(k)&=\frac{1}{2}\qty[\epsilon_p-\epsilon_s-\omega_0+2(\tilde{t}_p-t_s)\cos k],\nonumber\\
  h_{F,0}^{(1)}(k)&=\frac{1}{2}\qty[\epsilon_p+\epsilon_s-\omega_0+2(\tilde{t}_p+t_s)\cos k].
\end{align}
Here $\sigma_i$ are Pauli matrices and $\sigma_0$ is the identity matrix. The Floquet generator $H_F^{(1)}(k)$ reproduces the stroboscopic evolution at integer multiples of the drive period $T=2\pi/\omega_0$. The corresponding BIS pair consists of the two quasimomenta $k_{L/R}$ satisfying $h_{F,z}^{(1)}(k_{L/R})=0$, while the quasienergy gap in this channel is opened by the transverse term $h_{F,y}^{(1)}(k)$. In the present basis choice, BISs are conveniently identified by the zeros of $h_{F,z}^{(\ell)}(k)$ in the effective Floquet Hamiltonian, and the associated topological invariant is encoded in the winding of the transverse field along the BIS. In our system, $h_{F,y}^{(\ell)}(k)$ typically satisfies $h_{F,y}^{(\ell)}(-k)=-h_{F,y}^{(\ell)}(k)$, establishing a one-to-one correspondence between each BIS pair and the associated topological invariant, which simplifies the identification of the topology~\cite{Zhang2020,Wang2024}. The total invariant for the Floquet band is obtained from the numerically calculated Floquet Hamiltonian, with details given in the Supplemental Material~\cite{SM}.

The existence of a nontrivial one-dimensional Floquet phase still relies on symmetry protection. Even when the modulation satisfies the even-time condition $H(-t+t_\phi)=H(t+t_\phi)$ with $t_\phi=\phi_0/\omega_0$, a nonzero $h_{F,0}(k)$ breaks conventional chiral symmetry. The realized phase nevertheless remains protected by a magnetic group symmetry together with a nonlocal chiral symmetry~\cite{Song2018}. Unless stated otherwise, we work in the even-time driving convention.

In the weak-modulation regime, the Floquet bands can be viewed as replicas of the static bands of $H_0$, shifted by $n\omega_0$ and folded into the quasienergy FBZ $E\in[-\omega_0/2,\omega_0/2]$, up to the additional $k$-dependent shift $h_{F,0}(k)$. As shown in Fig.~\ref{fig1}(c,d), single- and two-harmonic driving generate one and two band crossings within the FBZ, respectively, and the transverse coupling opens gaps at these points. Since the static Hamiltonian $H_0$ contains no interband ($s$-$p$) coupling, all BISs are drive induced, so one cannot prepare a corresponding well-polarized initial equilibrium state on the BIS. Previous BIS detection protocols, which rely on such an initial state to extract time-averaged spin textures along a given pseudospin axis~\cite{Zhang2020,Zhang2023}, are therefore not directly applicable here. This motivates the Ramsey interferometric scheme used below.

\begin{figure*}[ht]
\includegraphics[scale=0.26]{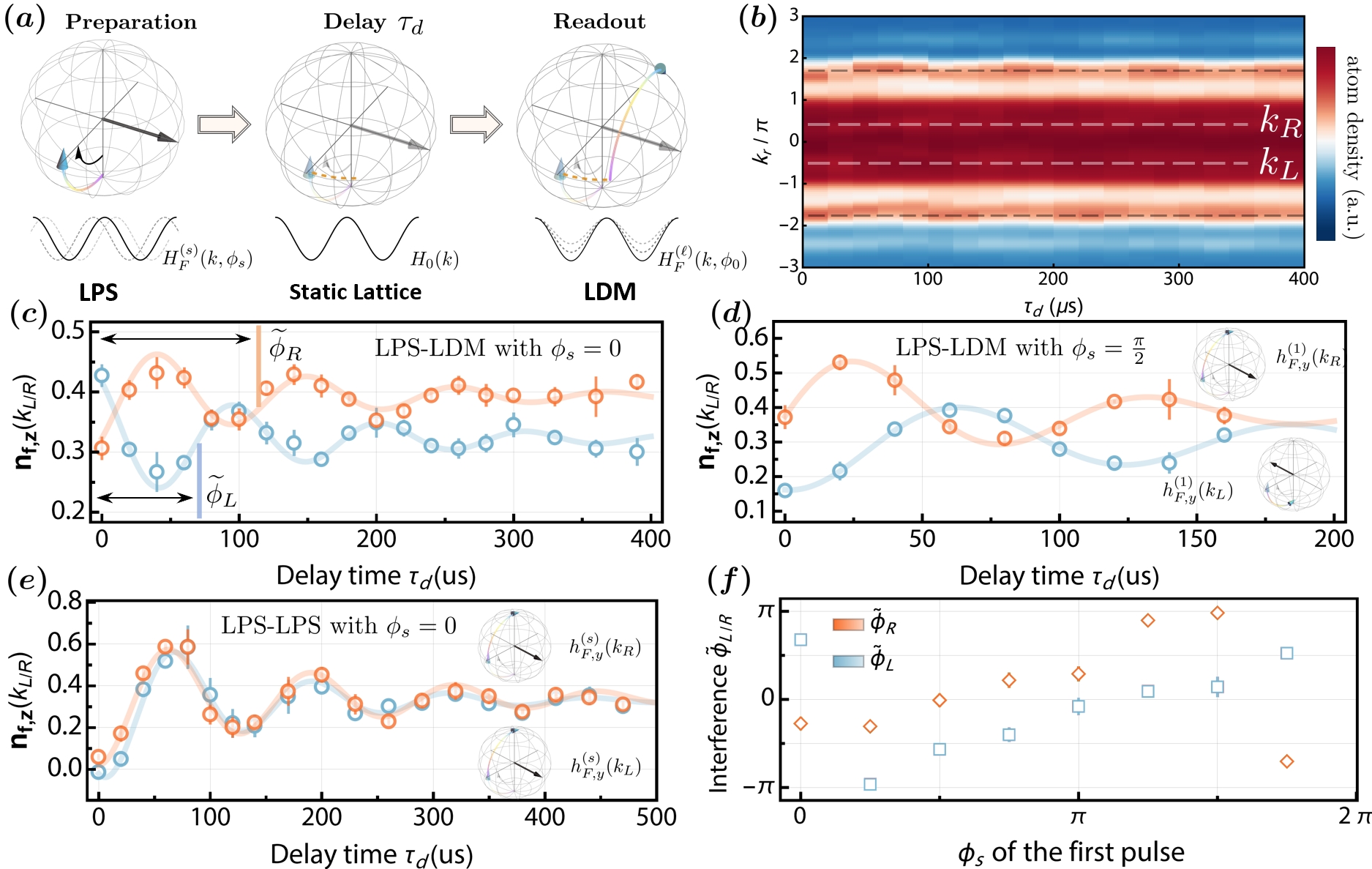}
\caption{\label{fig2}Ramsey detection of the single-photon $s$-$p$ resonance induced by single-tone lattice-depth modulation (LDM), measured at $k_L$ and $k_R$.
(a) The Ramsey sequence uses a lattice-position shaking (LPS) preparation pulse with stroboscopic form $H_F^{(s)}(k,\phi_s)$ and duration $t_1$, followed by a dark evolution of duration $\tau_d$ under the static Hamiltonian $H_0$, and a LDM readout described by $H_F^{(\ell)}(k,\phi_0=0)$ with duration $t_2$. All pulses are set close to $\pi/2$ to maximize fringe contrast.
(b) Experimental momentum-space density of the atoms measured by TOF imaging after the sequence. 
(c,d) Ramsey fringes of the spin imbalance at $k=k_R$ (yellow dots) and $k=k_L$ (blue dots) for the LPS--LDM sequence as $\tau_d$ is varied. The preparation phase choices are $\phi^{(s)}=0$ and $\pi/2$. A robust $\pi$ phase shift between $k_R$ and $k_L$ is observed.
(e) LPS--LPS sequence showing no phase shift between the two BIS fringes.
(f) In the LPS--LDM sequence, the LPS preparation phase $\phi_s$ sets the global fringe phase, while the relative $\pi$ contrast between $k_R$ and $k_L$ is preserved. The fitting function for the Ramsey fringe, shown as a solid line, is given by $n_{f,z}(t) = \sin(2\pi t/\tilde{T} + \tilde{\phi})e^{-\tilde{\lambda} t} + \tilde{c}$, where tildes denote fitted parameters.
The experimental parameters are $\delta D=22.3\,\mathrm{nm}$ for the LPS with $t_1=2T$ and $\delta V_0=1.2\,E_r$ for the LDM with $t_2=4T$. The drive period is $T=2\pi/\omega_0$ with $\omega_0=2\pi\times 8\,\mathrm{kHz}$ and $V_0=4.3\,E_r$ for static lattice.}
\end{figure*}

The Ramsey interferometric measurement is implemented by the mixed-modulation sequence shown in Fig.~\ref{fig2}(a): (i) $^{87}\mathrm{Rb}$ atoms are adiabatically loaded into the $s$ band, preparing an initial distribution that nearly fills the band at a temperature of about $105\,{\rm nK}$; (ii) a first-order LPS pulse of duration $t_1$ is abruptly applied as a topologically trivial beam-splitting pulse to couple the $s$ and $p$ bands, following the method first demonstrated in Ref.~\cite{kangCreutzLadderResonantly2020}; (iii) the atoms evolve in the static lattice for a time $\tau_d$; and (iv) an LDM pulse of duration $t_2$ is applied as the interferometric readout pulse. The LPS is described by $H_s=\hat{p}^2/2m+V_0\cos[2k_L(x+x_s(t))]/2$, where $x_s(t)=\delta D\cos(\omega_s t+\phi_s)$ and $\delta D$ is the maximum displacement. This drive defines the stroboscopic Floquet Hamiltonian $H_F^{(s)}(k,\phi_s)=\sum_i h^{(s)}_{F,i}(k,\phi_s)\sigma_i+h^{(s)}_{F,0}(k)\sigma_0$, where the superscript $s$ denotes the LPS Floquet Hamiltonian. We characterize the interferometric signal by the spin imbalance $\langle\sigma_z(k)\rangle=[n_s(k)-n_p(k)]/[n_s(k)+n_p(k)]\equiv n_{f,z}(k)$.

For this interferometric sequence, the resulting spin imbalance takes the form~\cite{SM}
\begin{align}\label{eq_spinimbaevlution}
  n_{f,z}(k)=-\qty[\mathcal{A}(k)+\mathcal{V}(k)\cos[\Phi(k)-\Phi_0(k)]],
\end{align}
where $\mathcal{A}(k)$ and $\mathcal{V}(k)$ denote a momentum-dependent offset and fringe visibility, respectively. Since the topological invariant is read out on the BIS, we focus on the quasimomenta satisfying $h_{F,z}^{(\ell)}(k_{\ell,L/R})=0$. We further choose both pulses to be close to $\pi/2$ pulses by setting $t_1$ and $t_2$ to proper integer multiples of the driving period, which suppresses micromotion and ensures the validity of the stroboscopic description. The accumulated phase at $k_{\ell,L/R}$ is mainly given by $\Phi_0(k)=\phi_s-\arg[\bm{h}_F^{(\ell)}(k_{\ell,L/R})]+\omega_s t_1+(\epsilon_p-\epsilon_s)\tau_d$, while $\Phi(k)=2h_{F,z}^{(s)}(k_{\ell,L/R})\tau_d$ remains approximately zero in the weak-drive regime when the same modulation frequency $\omega_s=\omega_0$ is used. Here $\bm{h}_F^{(\ell)}$ denotes the effective Bloch vector associated with the Floquet Hamiltonian.

For single-tone driving, we drop the channel index and write $k_{L/R}\equiv k_{1,L/R}$, since only one BIS pair is involved. Fig.~\ref{fig2}(b) shows the Ramsey fringe evolution for this case. The measured spin imbalance in Fig.~\ref{fig2}(c) exhibits an approximately $\pi$ phase difference between the two BISs at $k_L$ and $k_R$, consistent with the sign reversal $h_{F,y}^{(1)}(k_L)=-h_{F,y}^{(1)}(k_R)$. From the fits, we obtain oscillation periods $\tilde{T}=107(2)\,{\rm \mu s}$ at $k_L$ and $109(3)\,{\rm \mu s}$ at $k_R$.


By contrast, we also apply two LPS pulses in the Ramsey sequence. In this case, the corresponding coupling is topologically trivial due to $h_{F,y}^{(s)}(k_L)=h_{F,y}^{(s)}(k_R)$, so the spin-imbalance evolution remains in phase, $\tilde{\phi}_L(\tau)=\tilde{\phi}_R(\tau)$, as shown in Fig.~\ref{fig2}(e). The fitted Ramsey fringe period is $\tilde{T}=126(2)\,{\rm \mu s}$, which is closer to the shaking period $2\pi/\omega_0$.


In practice, the momentum signals generated by the LPS and LDM pulses largely overlap, so we analyze the response near the dominant peak where the two cloud centers cannot be reliably separated. Empirically, the LPS--LDM sequence yields Ramsey fringes with a shorter period, whereas the LPS--LPS sequence shows a longer period close to the shaking period $(\tilde{T}\approx 2\pi/\omega_0)$. This difference is consistent with a small pulse-dependent shift of the BIS position caused by different higher-order Floquet corrections for the two pulse types. As a result, the selected quasimomentum is slightly detuned from exact resonance during the Ramsey sequence, leading to an additional phase accumulation. We further vary the initial phase $\phi_s$ from $0$ to $2\pi$ and find that the fitted initial phases $\tilde{\phi}_{R(L)}$ shift accordingly, while the relative $\pi$ phase difference between the two BISs of the same pair remains robust, as shown in Fig.~\ref{fig2}(f).

\begin{figure}[ht]
\includegraphics[scale=0.23]{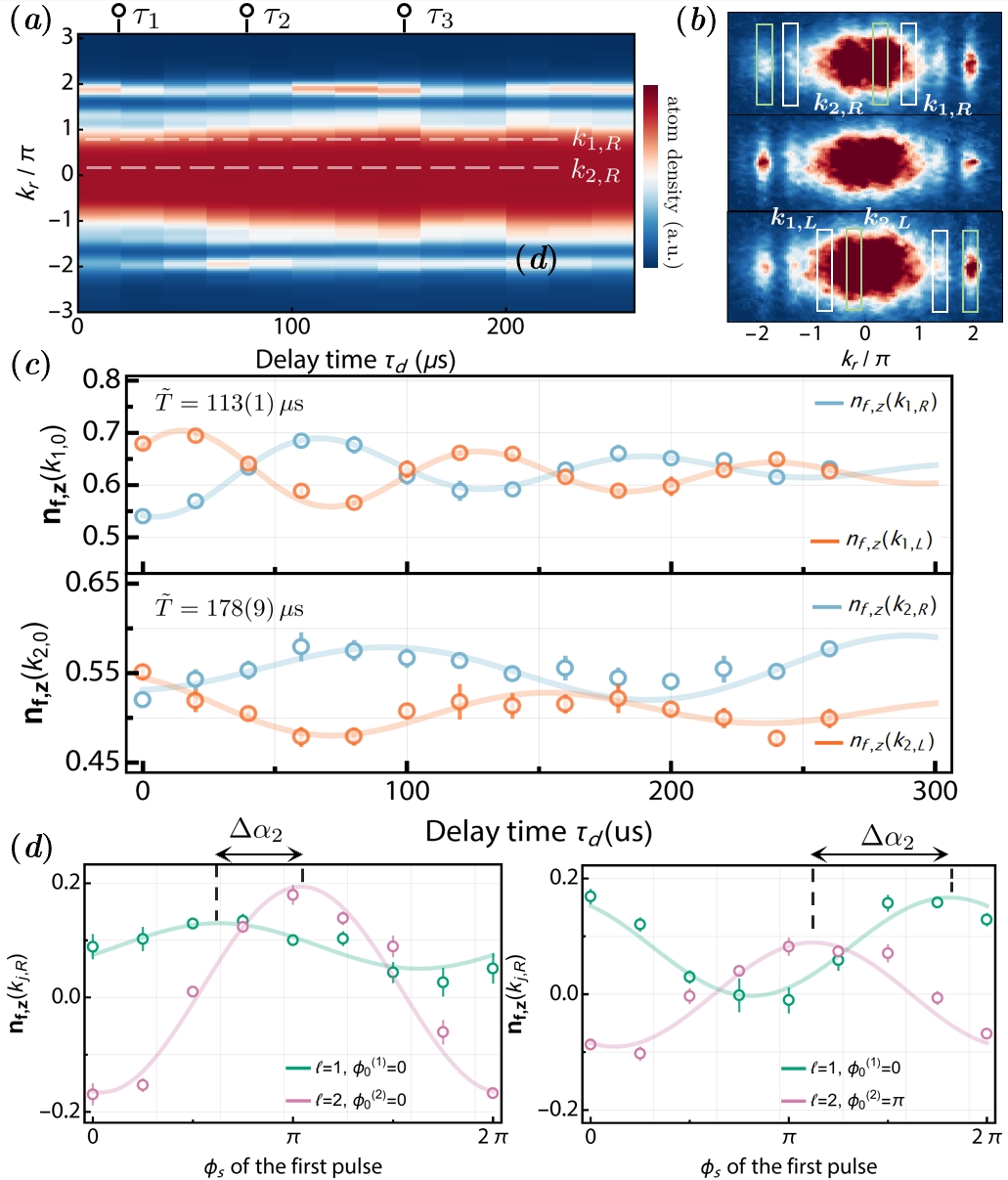}
\caption{\label{fig3}
Ramsey interferometry with two-tone driving $(\omega_0,2\omega_0)$.
(a) State preparation uses a two-tone LPS pulse, which launches coherent evolution governed by the Bloch equation of motion. Readout uses a two-tone LDM pulse with matching frequency components, so that both BIS pairs $(k_{\ell,L},k_{\ell,R})$ with $\ell=1,2$ are addressed. (b) Here $\tau_{1,2,3}$ label data taken at $\tau_d = 20\,\mu\mathrm{s}$, $80\,\mu\mathrm{s}$, and $160\,\mu\mathrm{s}$, respectively.
(c) A characteristic $\pi$ phase contrast between the two BISs of either the $0$ gap or the $\pi$ gap indicates opposite orientations of the effective Bloch field, as expected from $h_{F,y}(k_{\ell,L})=-h_{F,y}(k_{\ell,R})$.
(d) Zero-delay Ramsey readout for two settings of the read-out pulse phase, $\phi_0^{(2)}=0$ and $\pi$. Toggling $\phi_0^{(2)}$ produces a global $\pi$ shift of the interferometric phase and reverses the relative phase between the $\ell=1$ and $\ell=2$ BIS pairs. 
The experimental parameters are $\delta V_0^{(1)}=\delta V_0^{(2)}=0.98\,E_r$ for LDM, with a duration of $2T$. For the preparation LPS pulse, $\delta D^{(1)}=\delta D^{(2)}=22.3\,\mathrm{nm}$ and $\phi_s^{(1)}=\phi_s^{(2)}=\phi_s$, with a duration of $T$. The drive period is $T=2\pi/\omega_0$ with $\omega_0=2\pi\times 4\,\mathrm{kHz}$ and $V_0=3.5\,E_r$ for static lattice.}
\end{figure}

\begin{figure}[ht]
\includegraphics[scale=1.01]{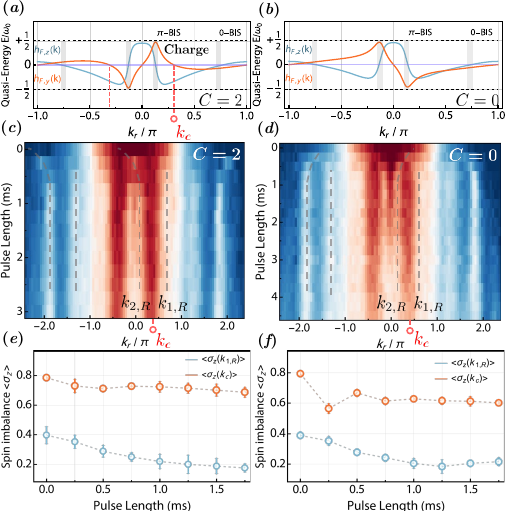}
\caption{\label{fig4} Measurement of the dynamics at the topological charge $k_c$, defined by $h_{F,y}(k_c)=0$. (a,b) Numerically calculated effective Floquet-band structures for $\phi_0^{(2)}=0$ and $\phi_0^{(2)}=\pi$, respectively. For $\phi_0^{(2)}=0$, the total winding number is $C=2$, and a topological charge appears at $k_c\approx 0.35\pi$ between the two BIS pairs. For $\phi_0^{(2)}=\pi$, the total winding number is $C=0$, and no topological charge lies between the two BIS pairs, so $\lvert h_{F,y}(k_c)\rvert$ remains finite. (c,d) Measured quench dynamics for $\phi_0^{(2)}=0$ and $\phi_0^{(2)}=\pi$, respectively. (e,f) Spin-imbalance evolution at $k_c$ extracted from (c,d). The topological charge at $k_c$ for $\phi_0^{(2)}=0$ leads to nearly frozen dynamics, whereas damped $s$-$p$ oscillations are observed for $\phi_0^{(2)}=\pi$. The dashed lines mark the momenta of interest and serve as guides to the eye. The color scale and the experimental LDM parameters are the same as in Fig.~\ref{fig3}. The calculated effective Floquet-Hamiltonian components $h_{F,y}(k)$ and $h_{F,z}(k)$ are obtained using the same parameters.}
\end{figure}

\section{Two-tone coherent modulation} \label{Sec3}


We next consider a two-tone LDM that simultaneously addresses the $\omega_0$ and $2\omega_0$ resonances through first-order couplings. The modulation is taken as
\begin{equation} \label{eq_twotoneV}
\delta V(t)=\delta V_0^{(1)}\cos(\omega_0 t+\phi_0^{(1)})+\delta V_0^{(2)}\cos(2\omega_0 t+\phi_0^{(2)}),
\end{equation}
with $\phi_0^{(1)}=0$ and $\phi_0^{(2)}=0$ or $\pi$ so as to preserve the even-time driving convention. The relative phase $\phi_0^{(2)}$ modifies the Floquet couplings and provides direct control over the topology of the driven bands. This mechanism differs from earlier two-tone LPS schemes, where the relative phase is mainly used to engineer gap closing at a selected BIS rather than to generate and control multiple BIS pairs in a gap-resolved manner~\cite{Kang2020,Minguzzi2022,Sandholzer2022}. Moreover, in an orbital-ladder setting, frequency-doubled two-tone LPS is not a clean route to anomalous Floquet topology, because the $2\omega_0$ component generically introduces a trivial first-order channel that masks the intended second-order topological coupling induced by the $\omega_0$ drive.

The two-tone drive in our scheme produces two drive-induced BIS pairs, namely the $0$-BIS pair ($\ell=1$) and the $\pi$-BIS pair ($\ell=2$). The former is described by $H_F^{(\ell=1)}(k)$ in Eqs.~\ref{Hamiltonian1} and \ref{Hamiltonian2}, while the latter is governed by a second Floquet Hamiltonian with the same Pauli-matrix decomposition as Eq.~\ref{Hamiltonian1}, but with different coefficients. The main terms are
\begin{align}
h_{F,y}^{(2)}(k) &= -\Big[\delta V_0^{(2)}\mathrm{e}^{-\dsi\phi_0^{(2)}} + \delta V_0^{(1)}J_1\!\big(\tfrac{\delta V_0^{(1)}}{\omega_0}\big)\Big]\nu_1^{ps}\sin k ,\nonumber\\
h_{F,z}^{(2)}(k) &= \tfrac{1}{2}\!\big[\epsilon_p-\epsilon_s-2\omega_0+2(\tilde{t}_p-t_s)\cos k\big],\nonumber\\
h_{F,0}^{(2)}(k) &= \tfrac{1}{2}\!\big[\epsilon_p+\epsilon_s-2\omega_0+2(\tilde{t}_p+t_s)\cos k\big],
\end{align}
where $J_n(z)$ is the Bessel function of the first kind. In the experimental regime, the direct second-harmonic term dominates, namely $\delta V_0^{(2)}\gg \delta V_0^{(1)}J_1(\delta V_0^{(1)}/\omega_0)$, so the gap at the BISs $k_{2,L/R}$ defined by $h_{F,z}^{(2)}(k_{2,L/R})=0$ is controlled primarily by the amplitude and phase of the $2\omega_0$ drive. For $\phi_0^{(2)}=0$ or $\pi$, the relative sign of $h_{F,y}^{(2)}$ with respect to $h_{F,y}^{(1)}$ is set by $\phi_0^{(2)}$, so the two BIS pairs contribute either additively or subtractively to the net winding. This phase control enables distinct configurations of $(W_0,W_\pi)$.

To probe the two BIS pairs experimentally, we employ the same Ramsey protocol as above. A two-tone LPS pulse with frequency components $\omega_0$ and $2\omega_0$, whose stroboscopic Floquet Hamiltonian is denoted by $H_F^{(s,\ell)}$, prepares Bloch states on the BIS pairs $k_{\ell,L/R}$ for $\ell=1,2$, and the second pulse is the two-tone LDM introduced in Eq.~\ref{eq_twotoneV}. By recording Ramsey fringes at the left and right BISs of each pair, we directly probe the relative phase between $h_{F,y}^{(\ell)}(k_{\ell,L})$ and $h_{F,y}^{(\ell)}(k_{\ell,R})$. As shown in Fig.~\ref{fig3}, the Ramsey fringes exhibit an approximately $\pi$ phase difference between $k_{\ell,L}$ and $k_{\ell,R}$ for both BIS pairs. The measured Ramsey periods are $\tilde{T}=178(9)\,\mathrm{\mu s}$ for the $\pi$-BIS and $\tilde{T}=113(1)\,\mathrm{\mu s}$ for the $0$-BIS. Because of the pulse-dependent period shift, the initial Ramsey phase does not directly yield the relative phase between the two BIS pairs, and therefore cannot by itself determine whether the two gap windings add or cancel.

We then examine the role of the second-tone phase $\phi_0^{(2)}$ using two consecutive pulses with zero inter-pulse delay $(\tau_d=0)$. The preparation pulse is a two-tone LPS pulse with initial phases $\phi_s^{(1)}=\phi_s^{(2)}\equiv\phi_s$ and a pulse area close to $\pi/2$, which places the Bloch vector near the equator on both BIS pairs. The readout is a two-tone LDM pulse that is shorter than a $\pi/2$ pulse and therefore acts only as a weak projection pulse, contributing little additional rotation while mainly imprinting the relative phase between the effective Bloch-field axes. In this limit, the phase contrast between the signals $n_{f,z}(k_\ell)$ measured on the $0$-BIS ($\ell=1$) and the $\pi$-BIS ($\ell=2$) is governed by $\Delta\alpha_2=\arg[h_{F,y}^{(2)}(k_{2,L/R})]-\arg[h_{F,y}^{(1)}(k_{1,L/R})]$. Fig.~\ref{fig3}(d) compares the cases $\phi_0^{(2)}=0$ and $\phi_0^{(2)}=\pi$ with $\phi_0^{(1)}=0$ fixed. The fitted values, $\Delta\alpha_2\approx 74^\circ$ and $-124^\circ$, differ by nearly $\pi$, showing that toggling $\phi_0^{(2)}$ reverses the orientation of the effective Bloch field $\bm{h}_F^{(2)}(k_{2,L/R})$ relative to $\bm{h}_F^{(1)}(k_{1,L/R})$. This demonstrates that the two BIS pairs are coherently linked by the same two-tone drive and hence jointly determine the global Floquet topology.

The phase reversal is directly reflected in the Floquet band topology. We numerically calculate the Floquet bands in the first Brillouin zone for $\phi_0^{(2)}=0$ and $\phi_0^{(2)}=\pi$ using a common gauge for the $0$- and $\pi$-gap sectors in Fig.~\ref{fig4}(a,b). For $\phi_0^{(2)}=0$, we obtain $(W_0,\,W_\pi)=(1,\,-1)$, giving a total winding $W=2$, whereas for $\phi_0^{(2)}=\pi$ we find $(W_0,\,W_\pi)=(1,\,1)$ and hence $W=0$~\cite{SM}. In both cases the topology is specified gap by gap through $(W_0,W_\pi)$, and the nonzero winding in the $\pi$ gap highlights the intrinsically Floquet character of the phase. Notably, when $\phi_0^{(2)}=\pi$, the net winding cancels while topological edge modes remain in the $0$ and $\pi$ gaps.

Away from the BISs, changing $\phi_0^{(2)}$ also reshapes the effective transverse field and could be extracted from the numerically calculated Floquet Hamiltonian. For $\phi_0^{(2)}=0$, $h_{F,y}(k)$ crosses zero at an intermediate quasimomentum $k=k_c$ between the two BIS pairs, reversing the sign of the effective coupling. This zero, defined by $h_{F,y}(k_c)=0$ with $h_{F,z}(k_c)\neq 0$, corresponds to a topological charge of the effective Bloch field~\cite{zhang2019}. We probe this feature by measuring the spin imbalance $\langle\sigma_z\rangle$ after a two-tone LDM pulse. Fig.~\ref{fig4}(c-f) compare the two phase choices. For $\phi_0^{(2)}=0$, a node appears near $k_c\approx 0.35\pi$, and the dynamics at that point are strongly suppressed, consistent with the vanishing coupling $h_{F,y}(k_c)=0$. For $\phi_0^{(2)}=\pi$, no such node appears between the BISs, and the dynamics near the midpoint instead show fast, damped oscillations between $s$ and $p$ orbitals. By contrast, the response near $k_{1,R}$ remains nearly unchanged in the two cases. This indicates that the observed contrast is tied specifically to the emergence of the $h_{F,y}=0$ node at $k_c$ for $\phi_0^{(2)}=0$, providing a dynamical signature of the phase-controlled restructuring of the Floquet bands beyond the BISs.

\section{Concluding Remarks}\label{Sec4}


In conclusion, we have realized a controllable, gap-resolved Floquet topological phase in a one-dimensional optical lattice. Beyond demonstrating multi-frequency control, we identify and exploit an important feature of LDM, namely the sizable nearest-neighbor $s$-$p$ orbital overlap. Because the $p$ orbital has odd parity, on-site $s$-$p$ coupling is absent by symmetry, while the nearest-neighbor overlap naturally gives rise the staggered sign structure required for a minimal nontrivial Floquet topology. A two-tone drive with tunable relative phase then controls the effective coupling signs in the $0$ and $\pi$ gaps separately. We read out the gap windings $(W_0,W_\pi)$ with a BIS-resolved Ramsey protocol, while controlled quenches reveal phase-dependent Floquet-band modifications even far from resonance.

These results have three implications. First, LDM remains topologically robust once realistic nearest-neighbor $s$-$p$ overlaps are included, establishing LDM as a quantitatively reliable knob for engineering orbital couplings and synthetic gauge structure in optical lattices~\cite{Li2013,Junemann2017,rajeevan2026}. Second, coherent multi-tone driving provides phase-tunable control of the sign pattern of the gap windings, while the BIS-resolved Ramsey protocol offers a compact interferometric readout performed gap by gap. Third, LPS and LDM, including multiple harmonics, can be combined phase coherently within a single sequence, enabling mixed-drive protocols with independently addressable gaps. These capabilities make multi-frequency phase control a practical tool for preparing and benchmarking a wide range of $(W_0,W_\pi)$ configurations and provide a practical route toward high-winding Floquet phases~\cite{Liu2019,Shi2024,Dalmin2024}. They also suggest that single-frequency step-wise protocols combining LPS and LDM may realize isolated anomalous $\pi$ modes in multiband systems~\cite{Ghuneim2025}.

{\it Acknowledgements.} We thank Long Zhang for fruitful discussions and for reading the manuscript. We also thank Zijie Zhu for helpful comments. This work was supported by National Key Research and Development Program of China (2021YFA1400900), the National Natural Science Foundation of China (Grants No. 12425401 and No. 12261160368), and the Quantum Science and Technology-National Science and Technology Major Project (Grant No. 2021ZD0302000). 

\noindent\textbf{Author contributions.}
P.J.Z. and Y.D.W. contributed equally to this work.

\bibliography{apssamp} 
\bibliographystyle{apsrev4-1} 
\clearpage

\onecolumngrid
\vspace*{1cm}
\begin{center}
	{\large\bfseries Supplementary Materials}
\end{center}
\setcounter{figure}{0}
\setcounter{equation}{0}
\renewcommand{\figurename}{Fig.}
\renewcommand{\thefigure}{S\arabic{figure}}
\renewcommand{\theequation}{S\arabic{equation}}

\section*{I.\ \ \ Effective Hamiltonian for multi-frequency lattice depth modulation} \label{AppendixA}

In this section we derive a general effective Hamiltonian for a one-dimensional lattice with multi-harmonic depth modulation (LDM). We start from the continuum Hamiltonian
\begin{align}
  H(t)&=\frac{\hat{p}^2}{2m}+\frac{1}{2}V_0\cos(2k_L x)+\frac{1}{2}V_{\rm mod}(t)\cos(2k_L x).
\end{align}
We assume the modulation has the form $V_{\rm mod}(t)=\sum_{j\geq 1}\delta V^{(j)}\cos(j\omega t+\phi_j)$ with $j$ being a positive integer. To preserve the even-time driving convention we set $\phi_j=0$ or $\pi$. Expanding the Hamiltonian in the $(s,p)$-orbital basis gives the multi-frequency two-band lattice model used in the main text,
\begin{align} \label{eq_realspaceH2}
  \hat{H}(t)=\sum_i\hat{\Psi}_i^\dagger M(t)\hat{\Psi}_i+\sum_i\qty[\hat{\Psi}_i^\dagger J(t)\hat{\Psi}_{i+1}+{\rm h.c.}],
\end{align}
with
\begin{align}
  M(t)&=\mqty(\epsilon_p  & 0 \\ 0 & \epsilon_s )+\mqty(\nu_0^{pp} & 0 \\ 0 & \nu_0^{ss} )V_{\rm mod}(t) 
\end{align}
and
\begin{align}
  J(t)&=\mqty(t_p & 0 \\0 & t_s )+\mqty(\nu_1^{pp} & \nu_1^{ps} \\ -\nu_1^{ps} & \nu_1^{ss} )V_{\rm mod}(t)
\end{align}
In momentum space the Hamiltonian becomes

\begin{align}
  \hat{H}(k,t)&=\mqty(\epsilon_p+2t_p\cos k & 0 \\ 0 & \epsilon_s+2t_s\cos k )+V_{\rm mod}(t)\mqty(\nu_0^{pp}+2\nu_1^{pp}\cos k & 2\dsi\nu_1^{ps}\sin k \\ -2\dsi\nu_1^{ps}\sin k & \nu_0^{ss} +2\nu_1^{ss}\cos k ) \nonumber \\
  &=\Delta_z(k)\sigma_z+\Delta_0(k)\sigma_0 +V_{\rm mod}(t) \qty[d_z(k)\sigma_z+d_0(k)\sigma_0+d_y(k)\sigma_y],
\end{align} 

where we have defined
\begin{align}
  \Delta_z(k)&=\frac{\epsilon_p-\epsilon_s+2(t_p-t_s)\cos k}{2}, \nonumber \\
  \Delta_0(k)&=\frac{\epsilon_p+\epsilon_s+2(t_p+t_s)\cos k}{2}, \nonumber \\
  d_z(k)&=\frac{\nu_0^{pp}-\nu_0^{ss}+2(\nu_1^{pp}-\nu_1^{ss})\cos k}{2}, \nonumber \\
  d_y(k)&=-2\nu_1^{ps}\sin k, \nonumber\\
  d_0(k)&=\frac{\nu_0^{pp}+\nu_0^{ss}+2(\nu_1^{pp}+\nu_1^{ss})\cos k}{2}.
\end{align}

We apply the time-dependent local unitary transformation
\begin{align}
  U&=\exp\qty[-\dsi \phi(t)\sigma_z/2], \nonumber \\
  \phi(t)&=2\Delta_ct+2d_z(k)\sum_{j\geq 1}\xi^{(j)} \sin (j\omega t+\phi_j),
\end{align}
with $\xi^{(j)}\equiv \delta V^{(j)}/(j\omega)$. The rotating-frame Hamiltonian reads $H_{\rm rot}=U\hat{H}(k,t)U^\dagger-\dsi U^\dagger\partial_tU $, then it takes the form
\begin{align}
  H_{\rm rot}
  &=\qty[\Delta_z(k)-\Delta_c]\sigma_z
  +\qty[\Delta_0(k)+d_0(k)V_{\rm mod}(t)]\sigma_0 +d_y(k)V_{\rm mod}(t)\qty[\cos\phi(t)\sigma_y+\sin\phi(t)\sigma_x] \nonumber\\
  &=\qty[\Delta_z(k)-\Delta_c]\sigma_z
  +\qty[\Delta_0(k)+d_0(k)V_{\rm mod}(t)]\sigma_0
  +\qty[-\dsi g(k,t)\sigma_+ +\dsi g^*(k,t)\sigma_-],
\end{align}
where $\Delta_c\equiv \ell \omega/2$ (with $\ell=\pm1,\pm2,\dots$) selects the resonant Floquet channel, and we defined $\sigma_\pm=(\sigma_x\pm \dsi\sigma_y)/2$.

The off-diagonal couplings are encoded in
\begin{align}
  g(k,t)\equiv d_y(k)V_{\rm mod}(t)\,{\rm e}^{\dsi\phi(t)}
  =\sum_m g_{m-\ell}(k)\,{\rm e}^{\dsi m\omega t}.
\end{align}
Here the integer indices $j$, $m$ (or $n$), and $\ell$ play diffe roles. The index $j$ labels the applied $j$th drive tone in the modulation $V_{\rm mod}(t)$, while $m$ and $n$ denote Fourier harmonic indices arising in the expansion of the dressed coupling $g(k,t)$. By contrast, $\ell$ labels the resonant Floquet channel, namely the $\ell$-photon resonance selected by the rotating-frame transformation through $\Delta_c=\ell\omega/2$.

The Fourier coefficients $g_n(k)$ can be obtained from the Jacobi-Anger expansion. Introducing $\beta^{(j)}\equiv 2d_z(k)\xi^{(j)}$, we write
\begin{align}
  {\rm e}^{\dsi \sum_{j\geq 1}\beta^{(j)}\sin(j \omega t+\phi_j)}
  =\sum_{n\in \mathbb{Z}}c_n(k)\,{\rm e}^{\dsi n\omega t},
\end{align}
with
\begin{align}
  c_n(k)=\sum_{\{n_j\}:\sum_j j n_j=n}\prod_{j\geq 1}
  J_{n_j}\qty(\beta^{(j)})\,{\rm e}^{\dsi n_j \phi_j},
\end{align}
where $J_n(z)$ is the Bessel function of the first kind. One then finds
\begin{align} \label{eq_Appgmk}
  g_n(k)=\frac{d_y(k)}{2}\sum_{j\geq 1}\delta V^{(j)}
  \qty[{\rm e}^{\dsi \phi_j}c_{n-j}(k)+{\rm e}^{-\dsi \phi_j}c_{n+j}(k)].
\end{align}

Expanding $H_{\rm rot}$ in Fourier components gives
\begin{align} \label{eq_AppHrot2}
  H_{\rm rot}
  &=\qty[\Delta_z(k)-\frac{\ell\omega}{2}]\sigma_z
  +\qty[\Delta_0(k)+d_0(k)V_{\rm mod}(t)]\sigma_0 \nonumber\\
  &\quad +\sum_m\qty[-\dsi g_{m-\ell}(k){\rm e}^{\dsi m\omega t}\sigma_+
  +\dsi g_{m-\ell}^*(k){\rm e}^{-\dsi m\omega t}\sigma_-].
\end{align}
Under the even-time driving convention ($\phi_j=0$ or $\pi$), the coefficients $c_n(k)$ and $g_n(k)$ are real. As a result, Eq.~\eqref{eq_AppHrot2} contains no $\sigma_x$ term (apart from the trivial $\sigma_0$ contribution), consistent with the conventional chiral-symmetry structure of the model. Moreover, the time average of the scalar modulation term vanishes:
\begin{align}
  \expval{d_0(k)V_{\rm mod}}_{T}
  =\frac{1}{T}\int_{0}^T d\tau\, d_0(k)V_{\rm mod}(\tau)=0
\end{align}
for even-time driving.

In the rotating frame, the resonant off-diagonal contribution is the static ($m=0$) Fourier component, namely $g_{m-\ell}(k)=g_{-\ell}(k)$. Retaining only this resonant term yields the effective Floquet Hamiltonian for the $\ell$th resonant channel,
\begin{align} \label{eq_SM15}
  H_F^{(\ell)}
  =\qty[\Delta_z(k)-\frac{\ell\omega}{2}]\sigma_z
  +\Delta_0(k)\sigma_0
  +\qty[-\dsi g_{-\ell}(k)\sigma_+ +\dsi g_{-\ell}^*(k)\sigma_-].
\end{align}
The remaining harmonics with $m\neq 0$ are off resonant and generate Bloch-Siegert shifts of the resonant sector. These corrections can be treated perturbatively. Since they are small compared with the modulation frequency and do not affect the topology, we neglect them in the present model.

\section*{II.\ \ \ Applications: single- and two-tone driving}
\begin{figure}[ht]
\includegraphics[scale=0.22]{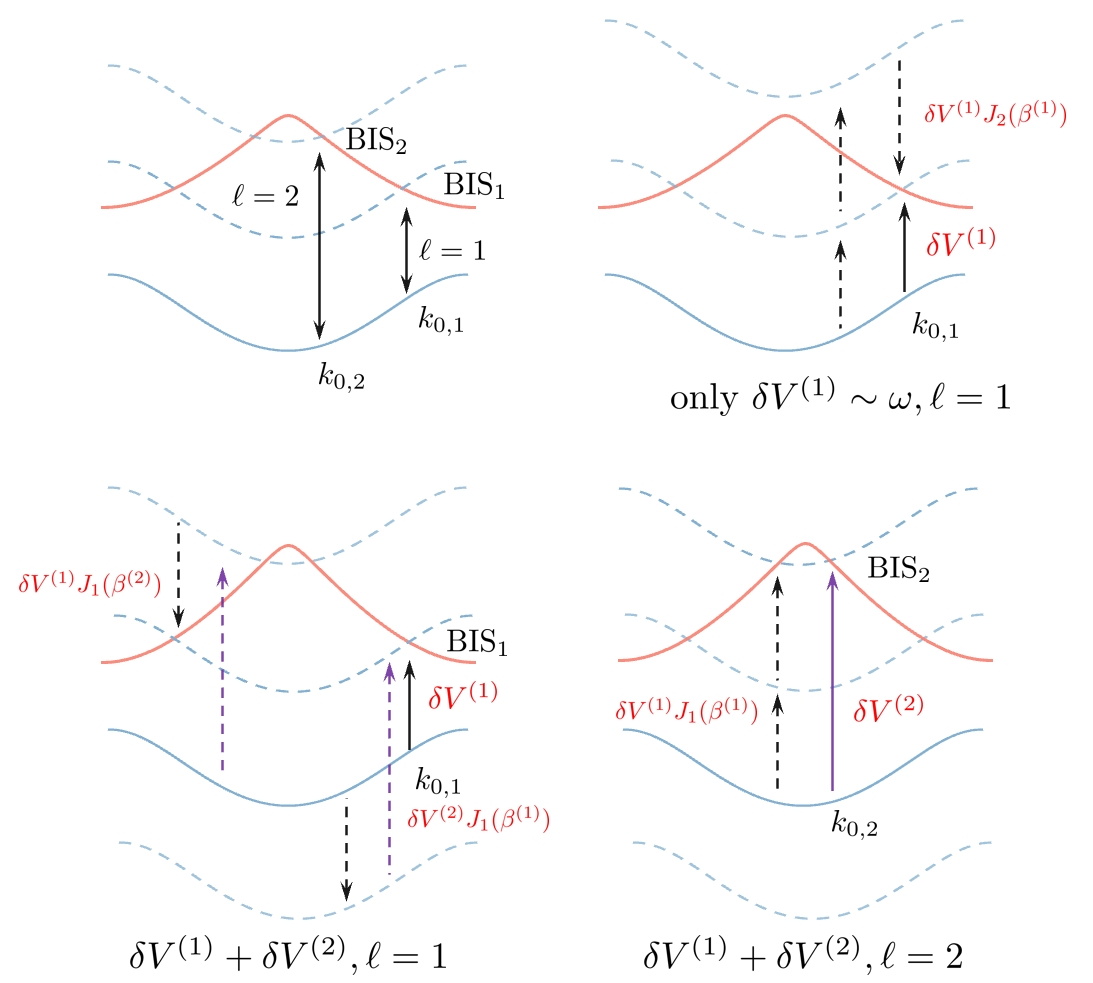}
\caption{\label{fig_SM1} 
Schematic illustration of the relevant resonant processes for single- and two-tone driving with multiple coherent channels. For two-tone driving, the single-photon resonance at the $\mathrm{BIS}_1$ is addressed by the first-order process $\delta V^{(1)}$ at frequency $\omega_0$, as well as by second-order processes involving combinations such as $\delta V^{(1)}J_1(\beta^{(2)})$ and $\delta V^{(2)}J_1(\beta^{(1)})$. The two-photon resonance at $\mathrm{BIS}_2$ is primarily driven by the first-order process $\delta V^{(2)}$ at $2\omega_0$, with additional contributions from higher-order processes involving $\delta V^{(1)}$.}
\end{figure}

For single-tone driving, where only the first drive tone $\delta V^{(1)}$ at frequency $\omega$ is present and $\phi_1=0$ for LDM, Eq.~\eqref{eq_Appgmk} gives
\begin{align} \label{eq_Appgmk2}
  g_{-\ell}(k)
  &=\frac{d_y(k)}{2}\,\delta V^{(1)}\big[ e^{\dsi\phi_1}c_{-\ell-1}(k)+e^{-\dsi\phi_1}c_{-\ell+1}(k)\big]
   =\frac{d_y(k)}{2}\,\delta V^{(1)}\frac{(-1)^{\ell+1}2\ell\,J_{\ell}\big(\beta^{(1)}\big)}{\beta^{(1)}}.
\end{align}
Here $\ell$ labels the resonant Floquet channel selected in the rotating frame. In the present single-tone case, where only the fundamental drive at frequency $\omega$ is applied, the $\ell$th channel also corresponds to a resonance involving a net energy transfer of $\ell\omega$. The corresponding coupling amplitude is governed by Bessel functions. In typical experimental regimes, the $s$-$p$ bandwidth is small compared with the bare orbital splitting, so only a single resonant channel, usually $\ell=1$ or $\ell=2$, is relevant.

For the $\ell=1$ resonant channel, Eq.~\eqref{eq_Appgmk2} reduces to
\begin{align}
  g_{-1}(k)
  &=\frac{d_y(k)}{2}\,\delta V^{(1)}\big[c_{-2}(k)+c_{0}(k)\big]\nonumber\\
  &=\frac{d_y(k)}{2}\,\delta V^{(1)}\qty[J_{2}(\beta^{(1)})+J_0(\beta^{(1)})]\nonumber\\
  &\approx\frac{d_y(k)}{2}\,\delta V^{(1)}\Big[1-\frac{(\beta^{(1)})^2}{8}\Big]\equiv h_{F,y}(k),
\end{align}
where the expansion is valid for weak driving, $|\beta^{(1)}|\ll1$. The term $c_{-2}(k)$ originates from the time-periodic modulation of $d_z(k)$ and represents a higher-order process that still contributes to the $\ell=1$ resonant channel through harmonic mixing. Retaining the $J_2(\beta^{(1)})$ contribution yields the estimate
\begin{align}
  \frac{d_y(k)}{2}\,\delta V^{(1)}c_{-2}(k)\simeq \frac{d_y(k)\,d_z(k)^2}{4\omega^2}\big(\delta V^{(1)}\big)^3,
\end{align}
consistent with a third-order amplitude scaling. Therefore the $\ell=1$ Hamiltonian takes the form
\begin{align}  \label{eq_SMsingletone}
  H_F^{(\ell=1)}(k)=\Big[\Delta_z(k)-\frac{\omega}{2}\Big]\sigma_z+\Delta_0(k)\sigma_0 + g_{-1}(k)\,\sigma_y .
\end{align}

We now consider two-tone LDM, where both the first and second drive tones $\delta V^{(1)}$ and $\delta V^{(2)}$ are present. For the $\ell=1$ resonant channel, the off-diagonal coefficient can be written as
\begin{align} 
  g_{-1}(k)
  &=\frac{d_y(k)}{2}\Big\{\delta V^{(1)}\big[c_{-2}(k)+c_0(k)\big]+\delta V^{(2)}\big[e^{\dsi\phi_2}c_{-3}(k)+e^{-\dsi\phi_2}c_1(k)\big]\Big\},
\end{align} 
where we have chosen the gauge $\phi_1=0$. Here the drive-tone indices refer to the applied modulation frequencies $\omega$ and $2\omega$, whereas $\ell=1$ specifies the resonant Floquet channel under consideration. For $\phi_2=0$ or $\pi$, the coefficients remain real. Introducing the generalized Bessel combination
\begin{align}
  J_n(z_1,z_2;\phi)\equiv\sum_{m\in\mathbb{Z}} J_{n-2m}(z_1)\,J_m(z_2)\,e^{\dsi m\phi},
\end{align}
and retaining only the leading contributions in the small-$\beta$ regime, one obtains
\begin{align}
  g_{-1}(k)
  &\approx\frac{d_y(k)}{2}\qty[\delta V^{(1)}\qty(c_{0}(k)+c_{-2}(k))+\delta V^{(2)}{\rm e}^{-\dsi\phi_2}c_1(k)] \nonumber\\
  &=\frac{d_y(k)}{2}\delta V^{(1)}\qty[J_0\qty(\beta^{(1)},\, \beta^{(2)}\,; \phi_2)+J_{-2}\qty(\beta^{(1)},\, \beta^{(2)}\,; \phi_2)]
   -\frac{d_y(k)}{2}\delta V^{(2)}{\rm e}^{-\dsi\phi_2}J_1\qty(\beta^{(1)},\, \beta^{(2)}\,; \phi_2) \nonumber\\
  &\approx\frac{d_y(k)}{2}\qty{\delta V^{(1)}+{\rm e}^{-\dsi\phi_2}\qty[\delta V^{(2)}J_1\qty(\beta^{(1)})-\delta V^{(1)}J_1\qty(\beta^{(2)})]}.
\end{align}
This expression shows explicitly that the second drive tone can coherently enhance or reduce the effective coupling in the $\ell=1$ channel, depending on its amplitude and relative phase $\phi_2$. In the weak-driving limit, $\beta^{(j)}\ll1$, the correction involving $J_1$ is small, and one recovers $g_{-1}(k)\approx d_y(k)\,\delta V^{(1)}/2$, namely the single-tone result in Eq.~\ref{eq_SMsingletone}.

For the $\ell=2$ resonant channel, and to the same order of approximation, we obtain
\begin{align}
  g_{-2}(k)
  &\approx \frac{d_y(k)}{2}\big[\delta V^{(1)}c_{-1}(k)+e^{-\dsi\phi_2}\delta V^{(2)}c_0(k)\big]\nonumber\\
  &\approx \frac{d_y(k)}{2}\Big[\delta V^{(2)}e^{-\dsi\phi_2}-\delta V^{(1)}J_1\big(\beta^{(1)}\big)\Big].
\end{align}
Here the $\ell=2$ channel corresponds to a total resonant energy transfer of $2\omega$. In the two-tone case, it receives a first-order contribution from the second harmonic and a second-order contribution from the fundamental tone. The corresponding effective Floquet Hamiltonian is therefore
\begin{align}
  H_F^{(\ell=2)}(k)=\Big[\Delta_z(k)-\omega\Big]\sigma_z+\Delta_0(k)\sigma_0 + g_{-2}(k)\,\sigma_y .
\end{align}

Thus, two-tone LDM allows different resonant pathways to interfere coherently within a given channel. The relative phase $\phi_2$ and the amplitudes $\delta V^{(1,2)}$ control the magnitude and sign of the effective transverse coupling $g_{-\ell}(k)$, thereby enabling the phase-tunable Floquet topological transitions discussed in the main text.

\section*{III.\ \ \ Determining topology from band inversion surfaces (BIS)}\label{AppendixC}

\begin{figure}[ht]
\includegraphics[scale=0.8]{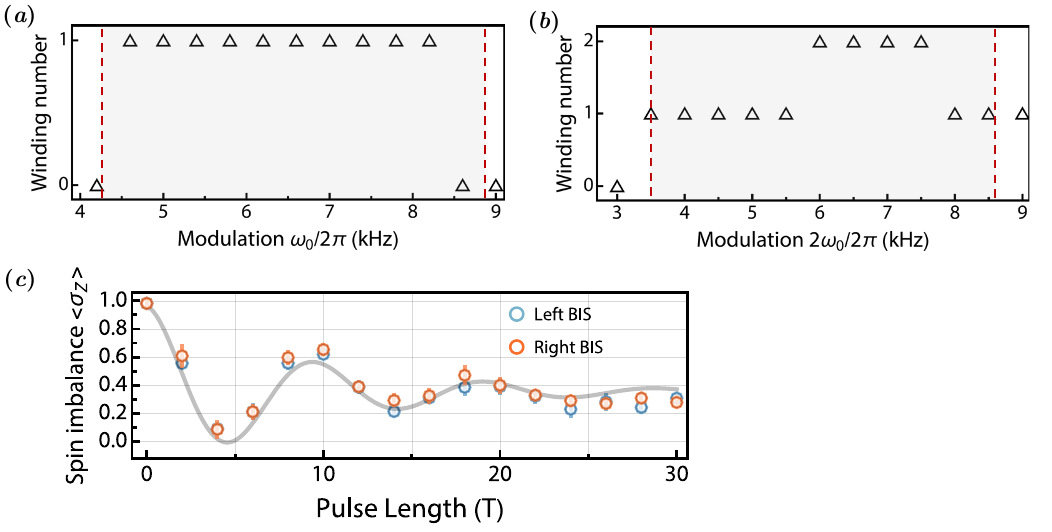}
\caption{\label{fig_SM3} (a,b) Winding number for Floquet bands with single-tone LDM at $\omega_0$ and two-tone LDM at $\omega_0$ and $2\omega_0$ in experiment. Parameters are identical to those in Fig.~\ref{fig2} for single-tone case and Fig.~\ref{fig3} for two-tone case of the main text. (c) Typical oscillations between $s$ and $p$ orbitals for single-tone LDM in experiments. Parameters for (c): $V_0=4.3\,E_r$, $\delta V_0=2\,E_r$, and modulation frequency $\omega_0/2\pi=8\,\mathrm{kHz}$.}
\end{figure}

We first present the winding number of the Floquet bands as a function of drive frequency, corresponding to the experimental results in Fig.~\ref{fig1}(c) of the main text, as shown in Fig.~\ref{fig_SM3}. For single-tone driving, the red dashed lines indicate the minimum ($\approx 4.24\,{\rm kHz}$ at $k=\pm\pi$) and maximum ($\approx 8.89\,{\rm kHz}$ at $k=0$) energy differences $\Delta_{sp}(k)=\qty[E_p(k)-E_s(k)]/h$ between the $s$ and $p$ bands for a static lattice with $V_0=4.3\,E_r$. For two-tone driving, the red dashed lines similarly indicate the minimum ($\approx 3.47\,{\rm kHz}$ at $k=\pm\pi$) and maximum ($\approx 8.62\,{\rm kHz}$ at $k=0$) values of $\Delta_{sp}$ for a static lattice with $V_0=3.5\,E_r$.

An interesting feature is that, in both cases, evident $s$-$p$ coupling already appears when the driving frequency $\omega_0$ ($2\omega_0$ in the two-tone case) lies slightly below the minimum single-photon resonance condition $\Delta_{sp}^{\min}$. This is consistent with the weak-driving picture, in which the periodic modulation can be viewed primarily as generating shifted replicas of the static lattice bands. As the drive frequency increases toward $\Delta_{sp}^{\max}$, however, the resonance window terminates slightly earlier in both cases. We attribute this mainly to coupling to higher bands, such as the $d$ orbital, which lowers the $p$-band energy near $k=0$ and thereby reduces the effective $s$-$p$ resonance window below $\Delta_{sp}^{\max}-\Delta_{sp}^{\min}$.

\begin{figure}[t]
\includegraphics[scale=1]{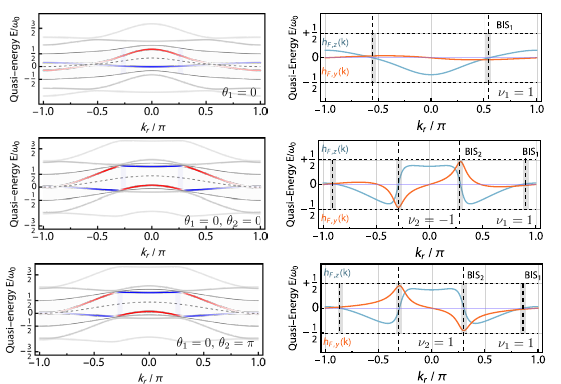}
\caption{\label{fig_SM2} Numerical identification of the BIS momenta and their contributions to the Floquet band winding number. The lattice depth is as in Fig.~\ref{fig3} of the main text, with driving amplitudes $\delta V = 0.5\,E_r$ (single-tone, $\omega_0=5\,E_r/\hbar$) and $\delta V_0^{(1)}=V_0^{(2)}=0.5\,E_r$ (two-tone, $(\omega_0\,,2\omega_0)$ with $\omega_0=3.5E_r/\hbar$).}
\end{figure}



In practice, the effect of the periodic drive is more intricate than the leading resonant picture alone, since it can generate not only the dominant transverse coupling but also additional effective terms, such as corrections along the $\sigma_z$ direction. We therefore directly compute the full stroboscopic Floquet Hamiltonian numerically. The time-evolution operator over one period is
\begin{align}
  U(T)&=\mathcal{T}\exp\Big[-\dsi\int_0^T H(k,t)\,dt\Big],\nonumber\\
  H_F(k)&=\frac{\dsi}{T}\log\big[U(T)\big],
\end{align}
where $\mathcal{T}$ denotes time ordering. The Floquet Hamiltonian is decomposed as
\begin{align}
  H_F(k)=\sum_i h_{F,i}(k)\sigma_i+h_{F,0}(k)\sigma_0.
\end{align}
The BIS pairs are then identified numerically from the zeros of $h_{F,z}(k)$. We denote the left and right BISs of the $a$th pair by $k_{a,L/R}$, where the index $a$ labels the BIS pairs obtained from the full numerical Floquet Hamiltonian. Evaluating the transverse field on these BISs, we define the associated topological invariant as
\begin{align}\label{eq_nu_a}
  \nu_a=-\frac{1}{2}\qty[{\rm sgn}\,h_{F,y}(k_{a,R})-{\rm sgn}\,h_{F,y}(k_{a,L})].
\end{align}
We adopt the same gauge for the $0$- and $\pi$-gap sectors, which enables direct comparison of the BIS contributions within the Floquet Brillouin zone. The gap-resolved winding numbers used in the main text are obtained by summing the BIS contributions associated with each quasienergy gap,
\begin{align}
  W_0=\sum_{a\in \mathcal{I}_0}\nu_a,\qquad
  W_\pi=\sum_{a\in \mathcal{I}_\pi}\nu_a,
\end{align}
where $\mathcal{I}_0$ and $\mathcal{I}_\pi$ denote the sets of BIS pairs belonging to the $0$ and $\pi$ gaps, respectively. The total winding of the lowest Floquet band in our convention is then
\begin{align}
  W=\sum_a (-1)^{\alpha_a/\pi}\nu_a,
\end{align}
where $\alpha_a$ takes the value $0$ or $\pi$ for a $0$-BIS pair or a $\pi$-BIS pair, respectively. In this convention, the total winding is determined by the gap-resolved invariants $(W_0,W_\pi)$ quoted in the main text.

Typical results are shown in Fig.~\ref{fig_SM2}. For single-tone LDM, only one BIS pair is found numerically, labeled by $a=1$. It gives $\nu_1=1$, and therefore $W_0=1$. For two-tone LDM, two BIS pairs are obtained and labeled by $a=1,2$. The relative phase of the second harmonic controls the sign of the effective transverse field on the second BIS pair and hence the value of $\nu_2$. Numerically, for $\phi_0^{(2)}=0$, we obtain $\nu_1=1$ and $\nu_2=-1$, corresponding to $(W_0,W_\pi)=(1,-1)$ and hence $W=2$. For $\phi_0^{(2)}=\pi$, we obtain $\nu_1=1$ and $\nu_2=1$, corresponding to $(W_0,W_\pi)=(1,1)$ and hence $W=0$.

On the other hand, the numerical results above show that the Floquet-band topology is governed primarily by the transverse coupling field, consistent with the effective Floquet-Hamiltonian picture. The topological properties can therefore also be understood from Eq.~\eqref{eq_SM15}. For the $\ell$th resonant channel, the corresponding BIS pair is identified from the resonance condition
\begin{align}
  \Delta_z(k_{\ell,0})-\frac{\ell\omega}{2}=0.
\end{align}
For single-tone driving, the effective Hamiltonian gives $h_{F,y}^{(1)}(k_{1,L})=-h_{F,y}^{(1)}(k_{1,R})$, and hence the BIS contribution of the $\ell=1$ channel is
\begin{align}
  \nu_1=-\frac{1}{2}\qty[{\rm sgn}\,h_{F,y}^{(1)}(k_{1,R})-{\rm sgn}\,h_{F,y}^{(1)}(k_{1,L})]=1.
\end{align}
For two-tone driving, the coupling in the $\ell=2$ channel is
\begin{align}
  h_{F,y}^{(2)}(k) = -\Big[\delta V_0^{(2)}\mathrm{e}^{-\dsi\phi_0^{(2)}} + \delta V_0^{(1)}J_1\!\big(\tfrac{\delta V_0^{(1)}}{\omega_0}\big)\Big]\nu_1^{ps}\sin k.
\end{align}
Its sign is controlled by the relative phase $\phi_0^{(2)}$ of the second harmonic with respect to the fundamental drive. Numerically, for $\phi_0^{(2)}=0$ we obtain $\nu_2=-1$. This extra minus sign can be understood from the inverted band ordering in the $\pi$-gap sector. Compared with the $\ell=1$ channel, the lower and upper Floquet bands are effectively exchanged, so the BIS contribution of the $\ell=2$ channel is reversed relative to the naive sign reading from $h_{F,y}^{(2)}(k)$ alone.

\section*{IV.\ \ \ General form for Floquet-Ramsey measurements}\label{AppendixD}
In this section we derive a general expression for the two-pulse Ramsey signal for different shaking protocols. From the Bloch-rotation viewpoint, the mapping from an initial Bloch vector $\bm{n}_0(k)$ to the final vector $\bm{n}_f(k)$ produced by two pulses separated by a dark time $\tau_d$ can be written as
\begin{align} \label{Appendix_D1}
  \bm{n}_f(k)=\hat{R}(\bm{n}_2,\Theta_2)\hat{R}(\bm{z},\Theta_z)\hat{R}(\bm{n}_1,\Theta_1)\bm{n}_0(k),
\end{align}
where $\hat{R}(\bm{n},\Theta)$ denotes rotation about axis $\bm{n}$ by angle $\Theta$, and $\Theta_i(k)=2\abs{\bm{h}_i(k)}t_i$ for the $i$-th pulse. The Rodrigues' rotation formula for arbitrary normalized vector $\bm{n}_0$ is
\begin{align}
  \hat{R}(\bm{n},\Theta)\bm{n}_0 = \bm{n}_0\cos\,\Theta+(\bm{n}\times \bm{n}_0)\sin\,\Theta+\bm{n}(\bm{n}\times\bm{n}_0)(1-\cos\,\Theta),
\end{align}
and we projecting onto the $z$-axis yields the general interferometric form for initial state $\bm{n}_0=(0,0,-1)$ in the main text
\begin{align} \label{eq_appendix_d3}
  n_{f,z}(k)=-\qty[\mathcal{A}(k)+\mathcal{V}(k)\cos[\Phi(k)-\Phi_0(k)]],
\end{align}
with
\begin{align}  \label{eq_appendix_d4}
  \mathcal{A}(k)&=\prod_{i=1}^2\qty[c_i+(1-c_i)(n_{i;z})^2], \nonumber\\
  \mathcal{C}(k)&=\prod_{i=1}^2\qty[n_{i;x}-(-1)^i \dsi n_{i;y}]\qty[(1-c_i)n_{i;z}+(-1)^i\dsi s_i], \nonumber \\
  \mathcal{V}(k)&=\abs{\mathcal{C}(k)},\qquad \Phi_0(k)=\arg{\mathcal{C}(k)}.
\end{align}
Here $c_i=\cos\Theta_i$, $s_i=\sin\Theta_i$, and $\bm{n}_i(k)=\bm{h}_i(k)/\abs{\bm{h}_i(k)}$ are the normalized rotation axes for pulses $i=1,2$. The dynamical phase accumulated during the dark interval is $\Phi(k)=2h_{1;z}(k)\tau_d$ if the first-pulse frame is used as frame reference.

As a concrete example, take the first pulse to be a  one-photon LPS resonance used to prepare BIS states. Its Floquet field is $\bm{h}_F^{(s)}(k)$ with components (see e.g.~\cite{kangCreutzLadderResonantly2020})
\begin{align}
  h^{(s)}_{F,x}(k)&=\mu_0^{sp}\cos\phi_s, \quad
  h^{(s)}_{F,y}(k)=\mu_0^{sp}\sin\phi_s, \nonumber \\
  h^{(s)}_{F,z}(k)&=\frac{1}{2}\qty[\epsilon_p-\epsilon_s-\omega_s+2(t_p-t_s)\cos k],
\end{align}
where \(\mu_0^{sp}\) is the site-shaking matrix element and \(\phi_s\) the shaking phase. Note that the spin-independent term $h^{(s)}_{F,0}(k)=-(t_p+t_s)\cos k$ dose not effect the rotation, hence we drop it in the above equations. The second pulse is taken to be the LDM readout described by \(H_F^{(j)}(k)=\sum_i h^{(j)}_{F,i}(k)\sigma_i\). Using Eqs.~\eqref{eq_appendix_d4} one obtains the general Ramsey-phase expression for a LPS--LDM joint sequence:
\begin{align}
  \Phi_0(k)&=\Delta\phi(k)+\arg\qty[(1-c_2)\frac{h_{2;z}(k)}{\abs{\bm{h}_{2}(k)}}+\dsi s_2] \nonumber \\
  &\quad -\arg\qty[(1-c_1)\frac{h_{1;z}(k)}{\abs{\bm{h}_{1}(k)}}+\dsi s_1], \nonumber \\
  \Delta\phi(k)&=\alpha_1-\alpha_2+\omega_1 t_1+(\epsilon_p-\epsilon_s)\tau_d,
\end{align}
where \(\alpha_i=\arg[h_{i;x}+\dsi h_{i;y}]\) and \(\omega_1\) is the carrier frequency of the first pulse. The fringe contrast is
\begin{align}
  \mathcal{V}(k)&=\frac{\abs{h_{F,\perp}^{(j)}(k)h_{F,\perp}^{(s)}(k)}}{\abs{\bm{h}_{F}^{(j)}(k)\bm{h}_{F}^{(s)}(k)}}\sqrt{\qty[(1-c_2)\frac{h_{F,z}^{(j)}(k)}{\abs{\bm{h}_{F}^{(j)}(k)}}]^2+s_2^2} \nonumber \\
  &\quad \times \sqrt{\qty[(1-c_1)\frac{h_{F,z}^{(s)}(k)}{\abs{\bm{h}_{F}^{(s)}(k)}}]^2+s_1^2},
\end{align}
with \(h_{F,\perp}=\sqrt{h_{F,x}^2+h_{F,y}^2}\). These expressions summarize the Ramsey response for the protocols used in this work and clarify the role of pulse angles, phases and BIS geometry in interpreting the measured fringes.


\nocite{*}

\end{document}